# Strong interaction regime of the nonlinear Landau-Zener problem for photo- and magneto-association of cold atoms

R. Sokhoyan[1,2], H. Azizbekyan[1], C. Leroy[2], and A. Ishkhanyan[1]

[1]*Institute for Physical Research NAS of Armenia, 0203 Ashtarak-2, Armenia*
[2]*Institut Carnot de Bourgogne, UMR 5209 CNRS, Université de Bourgogne, BP 47870, 21078 Dijon, France*

E-mail: aishkhanyan@gmail.com, sruzan@gmail.com

**Abstract.** We discuss the strong interaction regime of the nonlinear Landau-Zener problem coming up at coherent photo- and magneto-association of ultracold atoms. We apply a variational approach to an exact third-order nonlinear differential equation for the molecular state probability and construct an accurate approximation describing the whole time dynamics of the coupled atom-molecular system. The resultant solution improves the accuracy of the previous approximation by A. Ishkhanyan et al. [J. Phys. A **39**, 14887 (2006)]. The obtained results reveal a remarkable observation that in the strong coupling limit the resonance crossing is mostly governed by the nonlinearity while the coherent atom-molecular oscillations coming up soon after the resonance has been crossed are principally of linear nature. This observation is supposed to be general for all the nonlinear quantum systems having the same generic quadratic nonlinearity, due to the basic attributes of the resonance crossing processes in such systems. The constructed approximation turns out to have a larger applicability range (than it was initially expected) covering the whole moderate coupling regime for which the proposed solution accurately describes all the main characteristics of the system's evolution except the amplitude of the coherent atom-molecule oscillation, which is rather overestimated.

## 1. Introduction

In contrast to atomic Bose-condensates [1, 2], achieving molecular ones via standard laser cooling techniques [3-5] is complicated, since the laser cooling freezes only the centre-of-mass motion of a quantum object. In the case of atoms this is sufficient. However, molecules possess rotational and vibrational degrees of freedom. Hence, to create ultracold molecules, different approaches should be employed. Currently, there are several approaches to this problem, among which the most widely used techniques are the optical laser photoassociation [6, 7] and magnetic Feshbach resonance [8, 9].

For theoretical discussion of the specific field configurations applied within these techniques, of particular interest is the Landau-Zener model of linear resonance crossing ($\delta_t(t) = 2\delta_0 t$) at constant field amplitude ($U(t) = U_0 = \text{const}$) [10, 11]. This is because, in order to achieve high conversion efficiency, one has to apply a level-crossing field configuration [12]. Then, as it is well appreciated, the Landau-Zener model inevitably comes up as a natural starting point for studying such models. For this reason, this model has been a subject of intensive investigations over the last years (see, e.g., [13-19]).

In the present paper we re-examine the strong interaction regime of the resonance crossing in nonlinear systems involving quadratic nonlinearities generic for all the bosonic field theories. We reveal a general property of such processes, namely, we show that the whole time dynamics of the transition process is effectively divided into two distinct regimes. We find that in the strong coupling limit, the time dynamics of atom-molecule conversion process consists of the crossing of resonance in an essentially nonlinear manner followed by atom-molecular coherent oscillations that are principally of linear nature. This separation of the two processes is rather unexpected because of generic mixing of the corresponding terms in the governing equations. The general nature of this observation is due to the basic attributes of the specific form of the quadratic nonlinearity involved.

Applying a variational approach to an exact third-order nonlinear differential equation obeyed by the molecular state probability, we develop an approximation that accurately describes the whole time dynamics of the coupled atom-molecular system in the strong coupling limit. The formulas improve the accuracy of the previous result [19] by providing the next approximation term. It turns out that the proposed approximation is applicable also for the intermediate regime of moderate coupling. In this regime, the solution accurately describes all the main characteristics of system's time dynamics except the amplitude of coherent atom-molecule oscillation occurring at the end of the association process.

## 2. Mathematical treatment

In the mean field two-mode approximation, both photoassociation and Feshbach resonance are described by a basic semiclassical time-dependent nonlinear two-state model [12, 20, 21]

$$i\frac{da_1}{dt} = U(t)e^{-i\delta(t)}\overline{a}_1 a_2,$$
$$i\frac{da_2}{dt} = \frac{U(t)}{2}e^{i\delta(t)}a_1 a_1,$$
(1)

where $a_1$ and $a_2$ are the probability amplitudes of atomic and molecular states, respectively, $\overline{a}_1$ denotes the complex conjugate of $a_1$. The real functions $U(t)$ and $\delta(t)$ are the characteristics of the applied field. When photoassociation terminology is used, $U(t)$ is referred to as the Rabi frequency of the laser field, and $\delta(t)$ is the frequency detuning modulation function for which the derivative, $\delta_t(t)$, is the detuning of the laser field frequency from that of transition from the atomic state to the molecular one.

We start our discussion with changing from set (1) to the equation for the molecular state probability $p = |a_2|^2$ [18, 19, 22], which we write in the following factorized form

$$\left(\frac{d}{dt} - \frac{1}{t}\right)\left[p_{tt} - \frac{\lambda}{2}(1 - 8p + 12p^2)\right] + 4t^2 p_t = 0 \qquad (2)$$

(hereafter, the alphabetical index denotes differentiation with respect to the mentioned variable). Here we have passed to the dimensionless time by applying the scaling $t \to t/\sqrt{\delta_0}$ and have introduced the conventional Landau-Zener parameter $\lambda = U_0^2/\delta_0$. Note that system (1) describes a lossless process, hence, the total number of particles is conserved: $|a_1|^2 + 2|a_2|^2 = \text{const} = 1$, and note also that this normalization relation is incorporated in Eq. (2). We suppose that the system starts from the all-atom state so the initial conditions read:

$$p(-\infty) = 0, \quad p_t(-\infty) = 0, \quad p_{tt}(-\infty) = 0. \qquad (3)$$

Since we consider the strong interaction regime we suppose that the Landau-Zener parameter is large (equivalently, the field intensity $U_0^2$ is large enough or the detuning sweep across the resonance is sufficiently slow, that is the sweep rate $\delta_0$ is small). Hence, the second term in the square brackets in Eq. (2) adopts, in general, large value. Since for large $t$ the last term of the equation also adopts large value, we suppose that the leading terms in Eq. (2) are the last two so that we neglect, for a while, the term $p_{tt}$ thus arriving at a limit nonlinear equation of the first order. This equation admits two trivial stationary solutions, $p = 1/2$ and $p = 1/6$, and a nontrivial one. Unfortunately, for the initial condition $p(-\infty) = 0$ the nontrivial solution diverges as $t \to +\infty$ [19], hence, it cannot be directly applied as a proper initial approximation. In Ref. [19], an appropriate initial approximation was constructed via combination of the nontrivial solution with the trivial one $p = 1/2$. Using the constructed function as a zero-order approximation, the nonadiabatic transition probability has been calculated and it appeared that the final transition probability is expressed as a power of the Landau-Zener parameter [14, 19] in contrast to the familiar exponential prediction of the linear theory [10, 11]. However, this approach is rather complicated and it does not provide a clear treatment of the time dynamics of the association process. Here we make a step forward proposing a much simpler treatment of the problem that gives comprehensive understanding of the whole time evolution of the system. To achieve this goal, we use an extended limit equation which differs from that used in Ref. [19] by a term of the form $A/t$, where $A$ is a constant which is supposed to be small compared with other involved terms in order not to change the leading asymptotes. Due to this modification of the limit equation, we manage to construct a simple two-term approximation that accurately describes the whole time dynamics of the system. Importantly, the constructed solution reveals the main characteristics of the process in a simple and natural manner.

The *extended* limit equation, involving an adjustable constant $A$, is written as

$$\left(\frac{d}{dt} - \frac{1}{t}\right)\left[-\frac{\lambda}{2}(1-8p+12p^2) + A\right] + 4t^2 p_t = 0. \quad (4)$$

This equation is integrated via transformation of the independent variable followed by interchange of the dependent and independent variables. This results in a polynomial equation of the fourth degree for the limit solution $p_0(t)$:

$$\frac{\lambda}{4t^2} = \frac{C_0 + p_0(p_0 - \beta_1)(p_0 - \beta_2)}{9(p_0 - \alpha_1)^2(p_0 - \alpha_2)^2}, \quad (5)$$

where $C_0$ is the integration constant and

$$\alpha_{1,2} = \frac{1}{3} \mp \frac{1}{6}\sqrt{1 + \frac{6A}{\lambda}}, \quad \beta_{1,2} = \frac{1}{2} \mp \sqrt{\frac{A}{2\lambda}}. \quad (6)$$

For the initial condition $p_0(-\infty) = 0$ it holds $C_0 = 0$. Note that at $A = 0$ the quartic equation (5) reduces to a quadratic one since in this case three of the four parameters $\alpha_{1,2}$, $\beta_{1,2}$ become equal, $\alpha_2 = \beta_1 = \beta_2 = 1/2$. The solution to this quadratic equation diverges at $t \to +\infty$. However, for a *positive* $A$ the solution to the quartic equation (5) defines a bounded, monotonically increasing function which tends to a finite value less than $1/2$ when $t \to +\infty$ (Fig.1). This solution has all the needed features to be used as an appropriate initial approximation for constructing a solution to the problem. It is thus understood that introduction of the parameter $A$ is, indeed, an essential point.

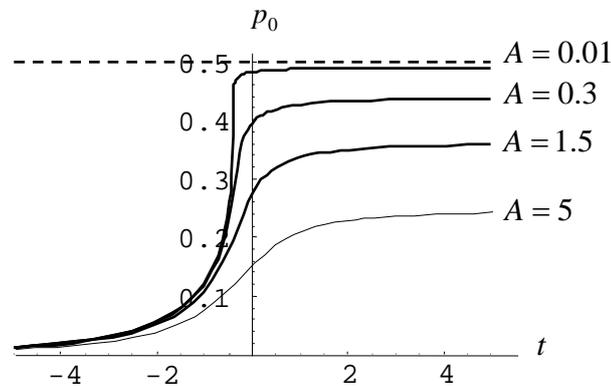

Fig.1. The limit solution $p_0(t)$ for positive $A > 0$ and a fixed $\lambda$.

Consider the properties of the limit solution $p_0(t)$ defined by Eq. (5) with $C_0 = 0$. The final value $p_0(+\infty)$ is easily found by noting that the left hand-side of Eq. (5) goes to zero as $t \to +\infty$. It is then seen that should be $p_0(+\infty) = 0$ or $p_0(+\infty) = \beta_1$ or $p_0(+\infty) = \beta_2$. Since $p_0(t)$ is a monotonically increasing function with $p_0(-\infty) = 0$ and since $\beta_2 > 1/2$, we deduce that $p_0(+\infty) = \beta_1$. In the similar way we find that $p_0(0) = \alpha_1$. Thus,

$$p_0(0) = \frac{1}{3} - \frac{1}{6}\sqrt{1 + \frac{6A}{\lambda}}, \quad p_0(+\infty) = \frac{1}{2} - \sqrt{\frac{A}{2\lambda}}. \tag{7}$$

To determine the appropriate value of the parameter $A$, we substitute $p_0(t, A)$ into the exact equation for the molecular state probability (2) and examine the remainder

$$R = \left(\frac{d}{dt} - \frac{1}{t}\right)[p_{0tt} - A]. \tag{8}$$

Obviously, the better the approximation $p_0$ is the smaller the remainder is. Now note that if $p_{0tt}(0) - A \neq 0$ the remainder diverges at the resonance crossing point $t = 0$ while it is finite for all other points of time. Therefore, we eliminate this divergence by requiring $A$ to obey the equation

$$p_{0tt}(0) - A = 0. \tag{9}$$

After some algebra, this equation is rewritten as

$$A = \frac{2}{9\lambda}\left(1 + \frac{1 - 18A/\lambda}{(1 + 6A/\lambda)^{3/2}}\right). \tag{10}$$

The approximate solution to the derived equation can be constructed by Newton's successive approximations starting, e.g., from $A = 0$. It turns out that the first approximation is already good enough. Thus, we put $A = 0$ in the right-hand side of the equation and obtain

$$A = \frac{4}{9\lambda}. \tag{11}$$

This value of $A$ leads to a good zero-order approximation $p_0(t)$. Numerical simulations show that for large $\lambda$ this function accurately describes the time evolution of the system in the interval covering the prehistory (up to the resonance point) and an interval after the resonance has been crossed. However, after that, $p_0$ misses several essential features of the process. Indeed, for instance, the coherent oscillations between atomic and molecular populations which come up at a certain time point after the resonance has been passed are not incorporated in this solution. Furthermore, the final transition probability at $t \to +\infty$ predicted by $p_0$ is always lower than what is shown by the numerical solution to the exact equation.

It is understood that the shortcomings of the suggested limit solution are due to the singularity of the procedure we have applied to obtain it. Indeed, we have constructed $p_0$ by neglecting the term $p_{tt}$ in the square brackets in Eq (2), i.e., the two highest order derivative terms of the equation. Of course, when determining the appropriate value of $A$ via imposing Eq. (9), we have taken into account these terms (in fact, to some extent). Yet, this was an indirect procedure and we have convinced that it is not enough.

Therefore, to improve the result, we need a correction that accounts for the second and third order derivatives of $p$. However, this is not an obvious task because the equation obeyed by the correction term $u \equiv p - p_0$ is still an essentially nonlinear one. Moreover, at an attempt to linearize the exact Eq. (2) using $p_0$ as a zero-order approximation and supposing the correction $u$ to be small as compared with $p_0$: $u \ll p_0$, we arrive at a complicated equation with variable coefficients (depending on $p_0$) the solution to which is not known. We now introduce an approach that enables one to overcome these difficulties. Importantly, the resultant solution not only correctly accounts for the higher order derivate terms in the equation for correction term $u$ but also takes into account, to a very good extent, the nonlinear terms. The constructed solution displays much more improved results. It both accurately treats the oscillations and well fits the final transition probability. For the most part of the variation range of the Landau-Zener parameter $\lambda \gg 1$, the resultant graphs are practically indistinguishable from the numerical solution.

Consider a correction $u$ defined as

$$p = p_0 + u. \tag{12}$$

This function obeys the following exact equation:

$$\left(\frac{d}{dt} - \frac{1}{t}\right)\left(u_{tt} + 4\lambda(1 - 3p_0)u + p_{0tt} - A - 6\lambda u^2\right) + 4t^2 u_t = 0. \tag{13}$$

Taking into account the initial conditions discussed here, we impose:

$$u(-\infty) = 0, \quad u_t(-\infty) = 0, \quad u_{tt}(-\infty) = 0. \tag{14}$$

Since the limit solution $p_0(t)$ is supposed to be a good approximation, the correction $u$ is expected to be small. So, we neglect, for a while, the nonlinear term $-6\lambda u^2$ in Eq. (13) thus arriving at a linear equation. Despite the fact that we now have a linear equation, there is only little progress since the solution to the derived equation in the general case of variable $p_0(t)$ is not known. However, note that in the case of a constant $p_0$ one can construct the solution using the scaling transformation

$$u = \frac{A}{2\lambda(1-3p_0)} v. \tag{15}$$

As a result, in this case we get a linear Landau–Zener problem for $v$ with an effective Landau–Zener parameter $\lambda^* = \lambda(1-3p_0)$.

This observation gives an argument to make a conjecture that the exact solution to Eq. (13) can be approximated as:

$$u = C^* \frac{p_{LZ}(\lambda^*, t)}{p_{LZ}(\lambda^*, \infty)}, \tag{16}$$

where $p_{LZ}(\lambda^*, t)$ is the solution to the linear Landau-Zener equation with an effective Landau-Zener parameter $\lambda^*$:

$$\left(\frac{d}{dt} - \frac{1}{t}\right)\left(p_{LZ\,tt} + 4\lambda^* p_{LZ} - 2\lambda^*\right) + 4t^2 p_{LZ\,t} = 0 \tag{17}$$

satisfying the initial conditions (3). This solution is conveniently written in terms of the Kummer hypergeometric functions [23] (see e.g., [18]).

This proves to be a good conjecture. The numerical simulations show that one can always find $C^*$, $\lambda^*$, and $A$ such that the approximate solution (16) accurately fits the numerical solution to Eq. (13).

Now, in order to derive analytic formulas for fitting parameters $C^*$ and $\lambda^*$, we substitute expression (16) into the exact Eq. (13) and aim at minimization of the remainder

$$R = \left(\frac{d}{dt} - \frac{1}{t}\right) \\ \left\{4[\lambda(1-3p_0) - \lambda^*]\frac{P_{LZ}(\lambda^*, t)}{P_{LZ}(\lambda^*, \infty)} + \frac{2\lambda^*}{P_{LZ}(\lambda^*, \infty)} + \frac{1}{C^*}(p_{0tt} - A) - 6\lambda C^* \frac{P_{LZ}^2(\lambda^*, t)}{P_{LZ}^2(\lambda^*, \infty)}\right\} \tag{18}$$

via appropriate choice of these parameters.

The first term in the curly brackets is a product of two functions. The function $\frac{P_{LZ}(\lambda^*, t)}{P_{LZ}(\lambda^*, \infty)}$ is an increasing (though oscillating) function that starts from zero at $t = -\infty$ and noticeably differs from zero only for time points $t > 0$. On the other hand, the function $4[\lambda(1-3p_0) - \lambda^*]$ is a monotonically decreasing function that tends to a large, since $\lambda$ is a large parameter, final value at $t \to +\infty$ (see Fig. 2). It is then understood that this term is highly suppressed if one chooses

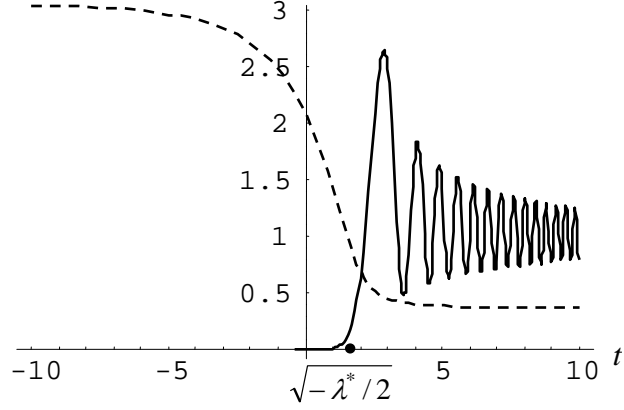

Fig. 2. Behavior of functions $4[\lambda(1-3p_0)-\lambda^*]$ (dashed line) and $P_{LZ}(\lambda^*,t)/P_{LZ}(\lambda^*,\infty)$.

$$\lambda^* = \lambda(1-3p_0(+\infty)). \tag{19}$$

Note that for $\lambda \gg 1$ this gives

$$\lambda^* \approx -\lambda/2 \tag{20}$$

so that for large $\lambda$, $\lambda^*$ becomes a large *negative* parameter. Interestingly, this choice of $\lambda^*$ leads to other relevant observations. First, it is known that

$$\lim_{t \to +\infty} P_{LZ}(\lambda^*, t) = 1 - e^{-\pi\lambda^*}, \tag{21}$$

hence, in the case of negative $\lambda^*$ the function $P_{LZ}(\lambda^*, \infty)$ grows exponentially with $|\lambda^*|$. Consequently, with this choice of $\lambda^*$ the second term in the curly brackets in Eq. (18) is also essentially suppressed. Second, in contrast to positive $\lambda^*$, for negative $\lambda^*$ the Landau-Zener function $P_{LZ}(\lambda^*, t)$ starts to noticeably differ from zero not merely for non-negative time points $t \geq 0$ but exclusively for those of the order of or larger than $\sqrt{-\lambda^*/2}$ (see Fig. 2). Hence, the first term in the curly brackets in Eq. (18) is even smaller than it was initially expected. Thus, the choice (19) essentially suppresses the first two terms in Eq. (18).

Regarding the two last terms in Eq. (18), one should minimize them with respect to the parameter $C^*$. This implies the condition

$$\frac{\partial R}{\partial C^*} = \left(\frac{d}{dt} - \frac{1}{t}\right)\left(-\frac{1}{C^{*2}}(p_{0tt} - A) - 6\lambda \frac{P_{LZ}^2(\lambda^*, t)}{P_{LZ}^2(\lambda^*, \infty)}\right) = 0. \tag{22}$$

Since the last term of this equation is proportional to (large) $\lambda$ and $\dfrac{P_{LZ}(\lambda^*,t)}{P_{LZ}(\lambda^*,\infty)}$ is an increasing function of time, it is understood that the "worst" point is $t = +\infty$. Hence, we look for minimization at $t = +\infty$. This immediately leads to the following value for $C^*$:

$$C^* = \sqrt{\frac{A}{6\lambda}}. \tag{23}$$

This result, together with relation (19), is of considerable general importance. Indeed, we see that though we use a solution to a linear equation, $P_{LZ}(\lambda^*,t)$, the parameters of this solution, $\lambda^*$ and $C^*$, are essentially changed due to the nonlinear terms involved.

The obtained formulas (19) and (23) present a rather good approximation. As it can be checked numerically, the solution (12), $p = p_0 + u$, with $p_0$ being the exact solution to the limit equation (4) and $u$ being a linear Landau-Zener function qualitatively well describes the process. This solution can be then used as an initial approximation for linearization of the initial equation (2).

However, more elaborate approaches can be suggested. An immediate observation, e.g., is that if we try the approximation (12), (16) without imposing the initial restriction that the introduced parameter $A$ is already determined by Eq. (9), one may modify the latter equation to determine a value of $A$ which will take into account the correction term $u$. The development of this approach leads to the following formulas for $\lambda^*$ and $C^*$:

$$\lambda^* = -\frac{\lambda}{2} + \lambda \cdot \ln\left(1 + \frac{1}{\lambda}\right), \tag{24}$$

$$C^* = \frac{1}{4\lambda} + \frac{1}{27\lambda^3}. \tag{25}$$

These formulas define a fairly good approximation. Indeed, starting already from $\lambda = 3$, the produced graphs (Fig. 3) are practically indistinguishable from the numerical solution of the exact Eqs. (1). The derived approximation notably improves the accuracy of the previous approximation of Ref. [19]. However, importantly, it is applicable far beyond the strong interaction limit and provides a sufficiently good description also for intermediate regime of moderate field intensities (or sweeping rates) down to $\lambda = 1$ and even slightly less ($0.95 < \lambda < 1$) (Fig. 4). Though in this regime the predicted amplitude of oscillations differs from that displayed by the numerical solution, it is seen from Fig. 4 that the approximation correctly describes many properties of the system's time evolution including the effective transition time, the final transition probability, and the period of atom-molecule oscillations.

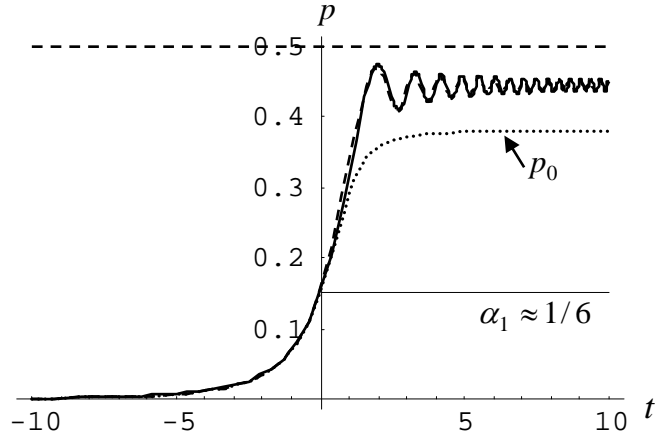

Fig. 3. Molecular state probability vs. time at $\lambda = 4$ (dashed line – approximate solution with parameters (11), (24), (25), dotted line – limit solution). It is seen that in the strong coupling limit $\lambda \gg 1$ the prehistory of the system and the resonance crossing are basically defined by the limit solution $p_0$ while the atom-molecule oscillations are described by the correction $u$.

This is, indeed, a rather unexpected result, especially, if one notes that at moderate coupling $\lambda \sim 1 \div 1.5$ the function $p_0(t)$ is very far from the exact solution, as it is seen from Fig. 4. An immediate conclusion following from this result is that it is not the limit solution $p_0$ that basically defines the time evolution of the system in this regime but the "correction" $u$ which was during our calculation envisaged to be small as compared with the limit solution.

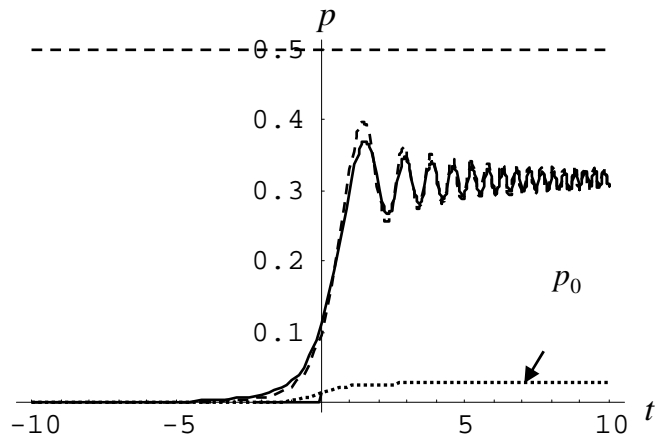

Fig. 4. Molecular state probability vs. time at $\lambda = 1$ (dashed line – approximate solution with parameters (11), (24), and (25), dotted line – limit solution). The limit solution $p_0$ is small so that it is the "correction" $u$ that basically defines the time evolution of the system in the regime of moderate coupling $\lambda \geq 1$.

Let us note in conclusion that the obtained formulas show that the final probability of the molecular state is given by the simple formula

$$p(+\infty) = \beta_1 + C^*. \quad (26)$$

Hence, the formula derived in Ref. [19] for the strong coupling limit $\lambda \gg 1$ is modified to include also the intermediate regime of moderate coupling $\lambda \geq 1$ as follows

$$p(+\infty) = \frac{1}{2} - \left(\frac{\sqrt{2}}{3} - \frac{1}{4}\right)\frac{1}{\lambda} + \frac{1}{27\lambda^3} \approx \frac{1}{2} - \frac{0.2214}{\lambda} + \frac{1}{27\lambda^3}. \quad (27)$$

Thus, for the quadratic nonlinear interaction we have discussed here the final probability for the system to stay in its initial all-atom state, $|a_1(+\infty)|^2 = 1 - 2p(+\infty)$, is not given by an exponential as predicted by the linear Landau-Zener theory [10, 11]. Instead, in the limit of strong coupling it is a linear function of the sweep rate $\delta_0 \sim 1/\lambda$ if the leading order of the approximation is discussed [14, 19]. This linear dependence of the non-transition probability on the sweep rate is confirmed to occur also by many-body calculations [15, 16, 17]. Note, finally, that formula (27) suggests the next approximation term as $1/(27\lambda^3)$.

**Conclusion**

We have presented an analysis of a quadratic-nonlinear version of the Landau–Zener problem that comes up in various physical situations, e.g., in photoassociation of an atomic Bose–Einstein condensate, in controlling the scattering length of an atomic condensate by means of Feshbach resonance, in second-harmonic generation, and generally in nonlinear field theories involving a Hamiltonian with a 2:1 resonance. Using an exact third-order nonlinear differential equation for the molecular state probability, we have developed an effective variational method for constructing the approximate solution to the problem in the strong coupling limit corresponding to the large values of the Landau-Zener parameter, $\lambda \gg 1$. In the case of photoassociation this implies that the intensity of the applied laser field is large enough or, equivalently, the sweep rate across the resonance is sufficiently slow.

We have shown that the approximation describing time evolution of the molecular state probability can be written as a sum of two distinct terms. In the strong coupling limit the first term, being a solution to a limit first-order *nonlinear* differential equation, effectively describes the process of the molecule formation while the second one, being the scaled solution to the *linear* Landau-Zener problem (but now with *negative* effective Landau-Zener parameter as long as the strong coupling limit of high field intensities or, equivalently, slow sweeping rates is considered), describes the oscillation which comes up some time after the system has passed through the resonance. From this, one can conclude that in the strong coupling limit the time dynamics of the atom-molecule conversion consists of the essentially nonlinear process of resonance crossing followed by atom-molecular coherent oscillations that are principally of linear nature. The possibility to make such a decomposition is quite surprising since the Hamiltonian of the system is essentially nonlinear.

The constructed approximation describes the molecule formation process with high accuracy. For $\lambda > 3$ the produced graphs are practically indistinguishable from the exact numerical solution (Fig. 3). Interestingly, the approximation rather well works also in the regime of moderate coupling down to $\lambda = 1$ (Fig. 4) and slightly less: $0.95 < \lambda < 1$. It correctly describes many properties of the system's time evolution including the effective transition time, the final transition probability, and the period of the atom-molecule oscillations. The only noticeable discrepancy is that the approximate solution overestimates the amplitude of the oscillations [the largest deviation is observed at the points of maxima and minima of the probability $p(t)$ within the time interval covering several first periods of oscillation]. The applicability of the proposed approximation to the intermediate regime of moderate coupling is, indeed, a rather unexpected result because at $\lambda \sim 1 \div 1.5$ the limit solution $p_0(t)$ is very far from the exact solution, hence, it is not the limit solution that mostly defines the evolution of the system in this regime. Using the constructed approximation one can easily find the main characteristics of the association process such as the tunneling time, the frequency of the oscillations of the transition probability that start soon after crossing the resonance, as well as the final transition probability to the molecular state. In particular, we have confirmed that the non-transition probability in the leading approximation order is a linear function of the sweep rate. In addition, we have found that the next approximation term is $1/(27\lambda^3)$.

Finally, we note that the presented approach is not restricted to the Landau-Zener model only. It can be generalized to other time-dependent level-crossing models [24, 25] too. Also, it can be adopted to explore other nonlinear regimes beyond by the Landau-Zener model [26]. Importantly, the developed approach allows one to treat the extended version of the nonlinear two-state state problem, when higher-order nonlinearities involving functions of the transition probability are added to the basic system (1). For example, one can analyze the role of the inter-particle elastic scattering which is described by Kerr-type cubic nonlinear terms [27]. Hence, the developed method may serve as a general strategy for attacking analogous nonlinear two-state problems involving the generic quadratic nonlinearity as discussed here.

**Acknowledgments**

This work was supported by the Armenian National Science and Education Fund (ANSEF Grant No. 2009-PS-1692) and the International Science and Technology Center (ISTC Grant N. A-1241). R. Sokhoyan and H. Azizbekyan acknowledge the French Embassy in Yerevan for the Grants No. 2006-4638 and No. 2007-3849 (Boursiers du Gouvernement Français). A. Ishkhanyan acknowledges Institut Carnot de l'Université de Bourgogne for the invited professorship in 2007 and 2009.

**References**

1. K.B. Davis, M.-O. Mewes, M.R. Andrews, N.J. van Druten, D.S. Durfee, D.M. Kurn, and W. Ketterle, Phys. Rev. Lett. **75**, 3969 (1995).
2. M.H. Anderson, J.R. Ensher, M.R. Matthews, C.E. Wieman, and E.A. Cornell, Science **269**, 198 (1995).
3. W.D. Phillips, Rev. Mod. Phys. **70**, 721 (1998).
4. C.N. Cohen-Tannoudji, Rev. Mod. Phys. **70**, 707 (1998).
5. S. Chu, Rev. Mod. Phys. **70**, 685 (1998).
6. A. Fioretti, D. Comparat, A. Crubellier, O. Dulieu, F. Masnou-Seeuws, and P. Pillet, Phys. Rev. Lett. **80**, 4402 (1998); A.N. Nikolov, E.E. Eyler, X.T. Wang, J. Li, H. Wang, W.C. Stwalley, and P.L. Gould, Phys. Rev. Lett. **82**, 703 (1999).
7. J. Weiner, V.S. Bagnato, S. Zilio, and P.S. Julienne, Rev. Mod. Phys. **71**, 1 (1999); F. Masnou-Seeuws, and P. Pillet, Adv. At. Mol. Opt. Phys. **47**, 53 (2001).
8. W.C. Stwalley, Phys. Rev. Lett. **37**, 1628 (1976); E. Tiesinga, B.J. Verhaar, and H.T.C. Stoof, Phys. Rev. A **47**, 4114 (1993).
9. S. Inouye, M.R. Andrews, J. Stenger, H.-J. Miesner, D.M. Stamper-Kurn, and W. Ketterle, Nature (London) **392**, 151 (1998).
10. L. D. Landau, Phys. Z. Sowjetunion **2**, 46 (1932); C. Zener, Proc. R. Soc. London, Ser. A **137**, 696 (1932).
11. E.C.G. Stuckelberg, Helv. Phys. Acta. **5**, 369 (1932); E. Majorana, Nuovo Cimento **9**, 45 (1932).
12. T. Kohler, K. Goral, and P.S. Julienne, Rev. Mod. Phys. **78**, 1311 (2006).
13. O. Zobay and B.M. Garraway, Phys. Rev. A **61**, 033603 (2000).
14. A. Ishkhanyan, M. Mackie, A. Carmichael, P.L. Gould, and J. Javanainen, Phys. Rev. A **69**, 043612 (2004).
15. I. Tikhonenkov, E. Pazy, Y.B. Band, M. Fleischhauer, and A. Vardi, Phys. Rev. A **73**, 043605 (2006).
16. E. Altman and A. Vishwanath, Phys. Rev. Lett. **95**, 110404 (2005).
17. R.A. Barankov and L.S. Levitov, cond-mat/0506323 (2005).
18. A. Ishkhanyan, J. Javanainen, and H. Nakamura, J. Phys. A **38**, 3505 (2005).
19. A. Ishkhanyan, J. Javanainen, and H. Nakamura, J. Phys. A **39**, 14887 (2006).
20. P.D. Drummond, K.V. Kheruntsyan, and H. He, Phys. Rev. Lett. **81**, 3055 (1998).
21. J. Javanainen and M. Mackie, Phys. Rev. A **59**, R3186 (1999); M. Koštrun, M. Mackie, R. Cote, and J. Javanainen, Phys. Rev. A **62**, 063616 (2000).
22. N. Sahakyan, H. Azizbekyan, H. Ishkhanyan, R. Sokhoyan, and A. Ishkhanyan, Laser Physics, **xx**, xxx (2009) (accepted).
23. M. Abramowitz and I.A. Stegun, *Handbook of Mathematical Functions* (Dover, New York, 1965).
24. E.E. Nikitin, Opt. Spectrosk. **13**, 761 (1962); E.E. Nikitin, Discuss. Faraday Soc. **33**, 14 (1962).
25. Yu.N. Demkov and M. Kunike, Vestn. Leningr. Univ. Fiz. Khim. **16**, 39 (1969).
26. A. Ishkhanyan, B. Joulakian, and K.-A. Suominen, Eur. Phys. J. D **48**, 397 (2008).
27. A. Ishkhanyan, R. Sokhoyan, K.-A. Suominen, C. Leroy, and H.-R. Jauslin, Eur. Phys. J. D **xx**, xxx (2009) (submitted).